\documentclass[aps, prd, twocolumn, lengthcheck, superscriptaddress, showpacs, letterpaper, nofootinbib]{revtex4-1}

\usepackage{graphicx}
\usepackage{color}

\def\la{\langle}\def\ra{\rangle}
\def\be{\begin{eqnarray}}\def\ee{\end{eqnarray}}
\def\lsim{\mathrel{\rlap{\lower3pt\hbox{\hskip1pt$\sim$}}
     \raise1pt\hbox{$<$}}} 
\def\gsim{\mathrel{\rlap{\lower3pt\hbox{\hskip1pt$\sim$}}
     \raise1pt\hbox{$>$}}} 
\def\le{ \begin{array}{ll}}\def\re{\end{array}}

\def\lear{ \left( \begin{array}{cc}}\def\rear{\end{array} \right)}

\def\le{ \left( \begin{array}{cc}}\def\re{\end{array} \right)}

\def\bi{\bibitem}

\begin{document}

\title{``Mass Without Mass" and Nuclear Matter}

\author{Mannque Rho}

\affiliation{%
Institut de Physique Th\'eorique,
CEA Saclay, 91191 Gif-sur-Yvette c\'edex, France
}

\date{\today}

\begin{abstract}
This is a brief account of the chief accomplishment of the  5-year activity (2008-2013) in the World Class University III Program at Hanyang University with the theme of exploring  ``From Dense Matter to Compact Stars." The principal objective was to explore and foresee what could be a break-through research that could be initiated at the RIB machine in project  ``RAON" at the upcoming ambitious Institute for Basic Science (IBS) in Korea. What came out was a possibility for unraveling  the mystery of proton's ``mass without mass" via nuclear matter and its impact on RAON-type physics and compact stars.

\end{abstract}


\maketitle

In 2008 the Korean Government launched what was called ``World Class University Program" with the objective to bring Korean Universities to the world's top level in basic science, and I was invited to join as a guest scholar in the program proposed by Hyun Kyu Lee -- and approved by the government -- at Hanyang University in the category III with supplementary support for workshops by the APCTP. The program III consisted of my spending four months a year for 5 years to help generate,  and participate in, original research activities at the graduate-school level in a field with potentials for a development of originality with an accent on ``future." The project chosen was what happens to baryonic matter under extreme gravitational pressure but stabilized against collapse into  black holes, i.e., massive compact stars, such as for instance the recently found $\sim 2$-solar mass objects~\cite{2solarmass}, and our objective was to explore what one can learn at the ``RAON" as a precursor to the phenomenon that takes place under extreme conditions. What transpired from the WCU/Hanyang (W/H for short) project is the glimpse into a possible source of the proton mass as revealed in nuclear processes. I dub the problem   ``mass without mass" borrowing the terminology from Wilczek's 1999 Physics Today article~\cite{wilczek}. It is indeed a prominent case of ``mass arising from nothing," a scenario of where the proton mass comes from and how it can manifest in nuclear processes that could be, hopefully, measured in RIB-type accelerators.

The landscape of hadronic phases, revealing the recent past in the cosmological evolution, has been fairly extensively explored at high temperatures thanks to lattice QCD on the theory side and to RHIC and now LHC on the experimental side. An amazingly elegant, and simple, structure  has been discovered in the form of highly correlated, nearly perfect liquid with quarks and gluons,  instead of weakly interacting quark-gluon plasma as naively thought in perturbative QCD. The situation of high density, relevant to massive compact stars, in stark contrast, is totally different.
The powerful theoretical tool, lattice QCD, cannot access the density relevant to the interior of compact stars, at present the only source available for physics at high density, and there are no other  tools that can be trusted, so the phase structure of baryonic matter beyond nuclear matter density is  totally barren and in the absence of experiments, poses a long-standing challenge in hadron/nuclear physics.

In this note I would like to describe the line of work that was performed at the W/H to go from nuclear matter density where nuclear matter is very well understood, to higher densities at low temperature. The principal theme was that the origin of the proton mass plays a key role in the equation of state (EoS) for compact stars, and can be explored in the forthcoming terrestrial accelerators such as RIB machines (e.g.,  RAON in Korea, FRIB of MSU/Michigan ...), FAIR of Darmstadt/Germany, NICA at Dubna/ Russia etc. and the space observatories in operation and in project. We set this as the principal objective of the W/H project, a promising direction for the coming era in hadron/nuclear physics in Korea.

The basic problem is that we do not know where more than 90\% of the proton mass, $\sim 1$ GeV (more precisely 938 MeV), come from. The mass of all the ``visible" objects around us, down to molecules and even to nuclei, can be accurately accounted for, say,   by more than 98\%,  by the total number of protons (and neutrons) included therein. When it comes to the mass of the proton it is no longer the sum of what constitutes its mass, namely quarks and gluons. The recent discovery of Higgs boson is to account for the mass of the quarks in the proton, but it does not ``explain" the proton mass: the quark masses are nearly zero, so whatever makes up the proton mass must be coming from something  other than ``sum of something."    In nature, there are even more fundamental mysteries such as neutrino masses, dark matter and so on.  In fact,  the proton mass is not really a {\it fundamental} issue, given that there is the theory QCD, its lattice calculation numerically giving nearly all of the proton mass: It comes out of ``back-reaction of gluon fields resisting accelerated motion of the quarks and gluons"~\cite{wilczek}. In the near future,  the QCD structure of the proton will be fully mapped out at the JLab in terms of quarks and gluons. But it will remain short of explaining why there are no ``stuffs" with mass that make up the proton mass.

The goal of the W/H project was to answer this question in terms of what we understand with nuclear matter and explore how to ``see" the mechanism in nuclear interactions. The objective and the early activity of this program were briefly recounted in \cite{LRS}. We have now arrived at what we think is the right answer, but it will still need confirmation and this is where the RAON/IBS can come in.

Let me state the problem in more physical terms. In the standard lore established by the pioneering work of Nambu, Goldstone and others, the proton mass (and the mass of  ``light-quark mesons,"  say, the $\rho$ meson to be specific), if one ignores  the tiny up and down quark masses, is said to be entirely ``generated dynamically."  Phrased in terms of symmetries, the mass is then attributed to the spontaneous breaking of chiral symmetry (SBCS for short). The SBCS is characterized by that the quark condensate $\la\bar{q}q\ra$ has non-vanishing vacuum expectation value (VEV), $\la\bar{q}q\ra_0\neq 0$. If this were the entire story, then one would have,  by `dialing' the quark condensate, that $
m_N(\la\bar{q}q\ra)\rightarrow 0 \ \ {\rm as}\ \ \la\bar{q}q\ra\rightarrow 0.$
But QCD does not say that this is the entire story. In fact, it is possible to have a mass term in the proton that does not vanish when the quark condensate goes to zero without violating chiral symmetry in the chiral limit. It is easy to understand this if one recalls the $SU(2)\times SU(2)$ Gell-Mann-L\'evy linear sigma model with the doublet nucleons, the triplet pions and the scalar $s$ meson (we reserve the usual notation $\sigma$ for the dilaton which will be introduced below). In this model, the nucleon and the scalar $s$ acquire masses entirely by the VEV of the $s$ field $\la s\ra_0\neq 0$ while the pion remains massless by Nambu-Goldstone theorem. On the other hand, as noted by DeTar and Kunihiro~\cite{detar}, there is nothing to prevent  the nucleon mass in an effective field theory from having the form
\be
m_N=m_0 +\bar{m} (\la\bar{q}q\ra)\label{pdoublet}
\ee
such that $\bar{m}\rightarrow 0$ as the condensate is dialed to zero,  {\em provided} one introduces parity doublet to the nucleon. Then the nucleon mass does not vanish if $m_0$ does not.  $m_0$ is a chiral-singlet.  What is significant is that  $m_0$ can be big, $m_0\approx (0.6-0.9)m_N$.

Now what about the mesons? In the framework developed at the W/H, the situation for meson masses is quite different. To address this matter,  we resort to an effective field theory that incorporates both scalar and vector degrees of freedom, the former as (pseudo-)Nambu-Goldstone (NG) boson, dilaton ($\sigma$), of spontaneously  {\it and} explicitly broken  -- via the trace anomaly -- symmetry and the latter $V=(\rho, \omega)$ as hidden local gauge fields. This approach naturally  brings the energy/momentum scale to the scale of meson mass,  $\sim 700$ MeV, in nuclear dynamics. The combined symmetry, scale-invariant hidden local symmetry ($s$HLS for short),  enables one to describe, at low orders in scale-chiral counting, baryonic matter up to the density commensurate with the interior of massive compact stars.

The scalar dilaton as a NG boson is highly controversial, because the existence of an infrared (IR) fixed point in QCD for small number of flavors, say, $N_F\sim 3$, is neither confirmed nor ruled out. Its existence  is strongly questioned in some circle of  particle theory community. It is also connected, for the number of flavors $N_F\gsim 8$, to the issue of dilatonic Higgs model with a view toward the Beyond  Standard Model. Our attitude is that whether the IR fixed point is present or not in QCD proper, its notion can be implemented  in nuclear processes at high density when combined with chiral symmetry, in the guise of scale-chiral perturbation theory. We find the scale symmetry, hidden or even absent in the vacuum, can ``emerge" as density increases at a dilaton-limit fixed point (DLFP) where the scalar $s$ becomes degenerate with the zero-mass pion in the chiral limit. This is in fact a notion associated with the dilaton condensate that has been around for a long time since the work I did with Gerry Brown, known as Brown-Rho scaling~\cite{BR91}\footnote{In retrospect, it should be mentioned that this scaling of 1991 applies only to density $\lsim 2 n_0$ and has been completely wrongly interpreted in the heavy-ion community with the outcome of dilepton experiments. The dilepton measurement at high temperature  does {\it not} probe this scaling associated with the dilaton condensate.}  and it reemerges in the work at the W/H but in a completely different guise.

As for the vector mesons,  it is hidden gauge symmetry that controls how their masses scale as density increases beyond the normal $n_0$.  While the flavor $U(2)$ symmetry for $V$ seems to be fairly good in the vacuum and also at low density in medium up to $n_0$, the symmetry is definitely broken at higher densities~\cite{u2symmetry}. So we focus on the $\rho$ meson which figures importantly at high density.
It has  been shown by a Wilsonian renormalization-group analysis that as the quark condensate goes to zero at chiral restoration, the $\rho$ meson mass should drop to zero as~\cite{HY:PR}
\be
m_\rho\sim g_\rho\sim \langle\bar{q}q\rangle \to 0.\label{rhomass}
\ee
This is the  ``vector manifestation (VM)" and the density at which the mass vanishes is the VM fixed point density.

Here then is a drastic difference between the proton mass  and the $\rho$ mass in the way chiral restoration is approached.  As density increases such that the quark condensate tends to zero, the ratio of the $\rho$ mass over the proton mass vanishes as
\be
m_\rho^\ast/m_N^\ast\propto \la\bar{q}q\ra^\ast/m_0\to 0.\label{massratio}
\ee
This is distinctly at odds with the ``Nambu scenario" based on spontaneous breaking of chiral symmetry, which would suggest the right-hand side of (\ref{massratio}) will be a constant $\propto 2/3$, the ratio of (constituent) quark numbers..

What we consider to be a pivotal development in the W/H project was the realization in 2010 that topology could play a valuable role in providing information on dense matter that is not available from elsewhere. The model that purports to unify, through topology, mesons and baryons in one Lagrangian formulated in 1959-1960 by Skyrme~\cite{skyrme} -- for nuclear physics -- was resuscitated in 1983 by field theorists (see  \cite{brown-zahed} for references) in terms of large $N_c$ QCD and quickly generated a great excitement among many particle and nuclear theorists. But the first string theory revolution -- which took place in 1984 -- took away most of the particle theorists, thereby leaving the field to nuclear theorists.
Unfortunately, due to daunting mathematical difficulties with the Skyrme model, not to mention the  skyrmion with $s$HLS  Lagrangian, and with many astute mathematical physicists gone to string theory,  the progress was extremely slow in formulating nuclear dynamics in the model.   Most of the practitioners of the model, highly frustrated with slow progress,  soon dropped the field and went over to simpler and easier work in chiral effective field theories that were being rapidly developed at about the same time.  A few die-hard skrymion aficionados  -- among whom I was one -- however stubbornly stuck to the model in attempting to solve one of the most challenging nuclear physics problems, dense baryonic matter. This effort was initiated in early 2000's at the newly established KIAS (Korea Institute of Advanced Study), and then was picked up at the W/H.  The discovery made at the W/H  resulted from this effort and led to the result that I describe below.

The progress made in the KIAS's period and after KIAS up to 2007 is reviewed by Byung-Yoon Park and Vicente Vento, the key actors in the effort, in \cite{multifacet}. Some of the works done during the W/H period are found in \cite{maetal} and reviewed by Harada, Lee, Ma and Rho  in \cite{multifacet}.

There was, and is even now, no  analytic or numerical skyrmion approach to dense matter with continuum Lagrangians. It is too difficult mathematically.  However dense matter can be simulated by putting skyrmions on crystal lattice. How to do this is described in various reviews and books (see for instance \cite{cndII, multifacet}). What turns out to come out of such simulations is new and crucially important for the physics of compact-star matter we were formulating. As the crystal size is reduced, corresponding to increased density, a skyrmion in medium is found to fractionize, at a density that we denote as $n_{1/2}$, into two half-skyrmions. This phenomenon is a topological effect, depending on symmetry,  and  is robust. This topological phenomenon resembles closely what takes place in some strongly correlated condensed matter systems such as fractional quantum Hall effects,  chiral superconductivity etc.~\cite{multifacet}. In our case with skyrmions,  it depends only on the pion degree of freedom that carries topology. The heavy degrees of freedom in $s$HLS do not affect its existence although they do influence  the precise location of the density involved, $n_{1/2}$.  Although it cannot be pinned down theoretically, phenomenology indicates that the reasonable location for $n_{1/2}$ is about $2n_0$. What follows in the calculations is insensitive to $n_{1/2}$ around $2n_0$. We have taken this as a typical value.

There is a good reason to believe that the crystal lattice approach can be justified at high density in the large $N_c$ limit of QCD, but it is most likely untrustworthy at low densities, i.e.,   near nuclear matter density $n_0$. Nonetheless the skyrmion-half-skyrmion transition itself, being topological, is very likely  to be robust.  We have taken this topological feature into account in our approach.

Now one of the most remarkable features of the half-skyrmion phase is that the quark condensate $\Sigma\equiv \la\bar{q}q\ra$, when averaged over the unit cell of crystal, $\hat{\Sigma}$, goes to zero while it is locally non-zero thereby supporting chiral density wave.  Chiral symmetry is not restored in this phase since the pion decay constant is nonzero. However the skyrmion representing the nucleon has a constant mass proportional to the dilaton condensate $\la\chi\ra$ where $\chi=f_\chi e^{\sigma/f_\chi}$, and $\sigma$ is the dilaton field.  The mass of the skyrmion can be written as
\be
m_{\rm skyrmion}=m^\prime_0+ \Delta (\hat\Sigma)\label{skyrmionmass}
\ee
and as $\hat{\Sigma}\to 0$, $\Delta$ goes to zero, so it reproduces Eq. (\ref{pdoublet}) with $m_0=m_0^\prime\propto \la\chi\ra$.  In the half-skyrmion phase, $\la\chi\ra$ is more or less constant with $m^\prime_0\sim (0.6 - 0.9)m_N$. This precisely reproduces the structure of the parity-doublet model. Note however the fundamental difference. Here the parity doubling emerges out of the medium whereas in the parity-doublet model, it is put in by hand.  This implies that even if QCD does not, intrinsically, have the parity-doublet symmetry, it can appear as an {\it emergent} symmetry. We argue that this is indeed the case and it is the scale symmetry that emerges.

Since a fully quantum-mechanical calculation in the skyrmion crystal description is not presently doable, the strategy adopted at the W/H was to map whatever robust features that could be extracted from it to an effective Lagrangian that has baryons put in explicitly as in standard nuclear chiral effective field theory. This can be done straightforwardly by coupling the baryons in a scale-chiral symmetric way  to $s$HLS with both scale and chiral symmetry breaking terms suitably incorporated. We call this Lagrangian $bs$HLS (``$b$" standing for the baryons and ``$s$" standing for the scalar (dilaton)). One can then set up a scale-chiral counting rule. This was done in \cite{CT-LMR}. Now the ``bare" parameters of the Lagrangian, matched to QCD at a suitable matching scale, carry QCD information, both perturbative and nonperturbative, that follow the vacuum change triggered by the dense medium. With this, one sets up Wilsonian renormalization-group (RG) effective field theory. How to do  RG effective field theory calculations given an effective Lagrangian like $bs$HLS  has been formulated elsewhere in various different ways. The technique we adopted, one of the most versatile and successful, is what's known as ``$V_{lowk}$ renormalization-group approach" ($V_{lowk}$-RG for short). The novelty in the W/H program was to provide the sliding-vacuum information in the Lagrangian $bs$HLS. It is in the latter that the topological inputs described above enter in a highly efficient way.

Several important impacts of the topology change come from  Eqs.~(\ref{rhomass}) and  (\ref{skyrmionmass}) with Eq.~(\ref{pdoublet}).  The first  is on the nuclear tensor force, one of the most important component of nuclear forces in nuclear interactions, which had been extensively studied since the early days of nuclear physics going back several decades. Now the topology change gives a totally new structure to it. In the $V_{lowk}$-RG with $bs$HLS, the tensor force is given entirely by the exchange of a pion and a $\rho$.   The important role of the pion has been well-known since Yukawa's discovery of the pion: It is what binds a proton and a neutron into deuteron, and figures prominently in complex nuclei, in particular in nuclei far from the stability line.   Equally significantly, the tensor force dominates in the symmetry energy, an essential part of the EoS for compact stars.  What's  new here however is that, firstly, the $\rho$ meson is indispensable for the tensor force -- this is being confirmed in on-going lattice QCD calculations of nuclear forces -- and, secondly, its tensor force is drastically quenched by the topology change with the $\rho NN$ coupling dropping rapidly as (\ref{rhomass}). Consequently, it undergoes a dramatic change at the density $n_{1/2}$ where the half-skyrmions appear. This observation made at the W/H in 2010 is described in ~\cite{LPR}. This phenomenon appeared there as a cusp at $n_{1/2}$ in the nuclear symmetry energy described in the skyrmion crystal.  This cusp may appear singularly non-normal, but turns out to have a very simple explanation~\cite{LPR} in terms of how the tensor force behaves as a function of density. The net tensor force, because of the cancelation between the two components, first decreases, as density increases, toward  $n_{1/2}$, and then increases afterwards due to the suppression of the hidden gauge coupling going toward the VM fixed point. In fact the decreasing tensor force as density approaches $n_0$ from below has been neatly confirmed by the long lifetime of the Carbon-14 (dating)~\cite{holt-c14}, and we expect -- and assume -- that this tendency continues up to $n_{1/2}$. What happens after $n_{1/2}$ as revealed in detailed calculations~\cite{dongetal} is that the effect of the topology change on the net tensor force makes the symmetry energy to stiffen from soft to hard at $n\sim 2n_0$. And this stiffening provides a simple mechanism for massive compact stars.

 Now what does this skyrmion-half-skyrmion cusp, having a qualitative effect on the nuclear tensor force, say about the source of the proton mass, nuclei far from stability  and massive compact stars?

The proton mass deduced from the skyrmion crystal, as noted above, goes like $m_N^\ast\propto \la\chi\ra^\ast$ in $n\gsim n_{1/2}\sim 2 n_0$. Here $\la\chi\ra$ is a chiral scalar, independent of density as in the parity-doublet model. As implied in (\ref{massratio}), the proton mass has a source quite different from the $\rho$ mass, the behavior of which is consistent with the Nambu scenario. Now to what density does the proton mass follow $\la\chi\ra$, staying independent of density? We will see later that this issue comes up with the sound velocity in massive stars.

There is something special and magical, hitherto unsuspected,  about the tensor force with its spin and isospin dependence, independently of where it comes from and also of the change as a function of density as manifested in the C-14 dating. There is a tantalizing hint, which could perhaps be explored in RAON-type experiments,  that the tensor force is {\it not} renormalized by strong interactions both inside and outside of nuclear medium. This surprising observation is described in \cite{pristine} with relevant references.  In Landau Fermi-liquid theory of nuclear matter,  a field theory way of understanding nuclear matter and also finite nuclei, the quasiparticle interactions are fixed-point quantities, becoming more accurate as $N\equiv k_F/\bar{\Lambda}$  -- where  $\bar{\Lambda}$ is the distance from the cutoff of the EFT to the Fermi surface -- goes to $\infty$.  The tensor force is not a local operator, so cannot be treated in the same way as one does for the Landau parameters. We have no proof that the tensor force is a fixed-point interaction, with a (nearly) vanishing $\beta$ function as suggested by the numerical analyses~\cite{pristine}. But there is a strong correlation with this tensor force and Landau-Migdal $G_0^\prime$ interaction which is at the fixed point as Gerry Brown argued (see \cite{pristine}). Let me suppose it is a fixed point interaction.  Then one can look at processes, such as the monopole matrix elements in exotic nuclei as studied by Otsuka~\cite{otsuka}, that could be studied in RAON-type accelerators probing the density regime $n\gsim 2 n_0$. That would reveal the ``dramatic change" of the tensor force as the density goes over $n_{1/2}$. Of course one would have to figure out what experiments can probe this property that zeroes in on the problem of proton mass vs. $\rho$ mass.  We have not had the time to work on this at the W/H but it could be a future project at the RAON/IBS.

Next what about the proton mass/$\rho$ mass issue with respect to compact stars? The answer to this question is given in the most recent publications by the post-W/H collaboration~\cite{PKLR,PR}. Once the density goes over $n_{1/2}$, the $\rho$ mass runs rapidly to zero, but the proton mass, with its {\it small} dynamically generated mass $\Delta (\Sigma)$ vanishing,  goes to a chiral-invariant constant proportional to the dilaton condensate $\la\chi\ra$. As mentioned, the dilaton condensate there is density-independent. Consequently the VEV of the trace of energy-momentum tensor, $\la\theta_\mu^\mu\ra$, becomes constant independent of density. Therefore $\partial \la\theta_\mu^\mu\ra/ {\partial n}= (\partial \epsilon (n)/\partial n)(1-3(v_s^2/{c^2}))=0$. This means that the sound velocity given by the EoS of the dense matter at $n > n_{1/2}$ is $v_s/c=\sqrt{1/3}$ because  $\frac{\partial \epsilon (n)}{\partial n}\neq 0$. This is a   surprising result.  Although there is, apparently,  no  known way to measure the sound velocity of dense matter in terrestrial laboratories,  it has a profound implication on the structure of compact-star matter. Theoretically  one would expect it to be $v_s/c < \sqrt{1/3}$ at low density, but driven by strong interactions between non-relativistic nucleons, increase first as density increases,  and then overshoot  $\sqrt{1/3}$ at $n > n_0$. Finally at asymptotically high density where perturbative QCD intervenes,  the sound velocity will asymptote to $\sqrt{1/3}$ from below due to the asymptotic freedom, with $\la\theta_\mu^\mu\ra\to 0$. In fact this is what is found in our theory. But in our theory, $n_{1/2}$ is merely a couple of times the normal matter density, potentially accessible at RIB accelerators and certainly in compact-star matter, so the ``conformality" seems to set in extremely precociously even though $\la\theta_\mu^\mu\ra\sim \la\chi\ra^4\neq 0$. How this surprising feature could appear in observables measurable  at densities at or exceeding $\sim 2 n_0$ is an extremely interesting question to address in the near future.  The RAON, perhaps, and surely the forthcoming accelerators such as FAIR etc. will be able to offer the glimpse into this totally uncharted region of strongly interacting matter. It could also be probed in gravity waves emitted from coalescing neutron stars to be measured in the future, for instance, in tidal deformability or Love number as predicted in the post-W/H collaboration~\cite{ PKLMR}.

We feel certain that what we have uncovered at the W/H is a big first step forward to ``seeing" what goes on at high density. But I must  say our effort has not been duly and fairly recognized, not to mention appreciated,  in the nuclear community, even in Korea.  When the discovery reported in \cite{LPR} of the cusp, the key ingredient of our development made at a pivotal moment, together with its implication on compact stars was submitted to Phys. Rev. Letter, the Divisional Associate Editor, after mediating serious differences between the authors and the referees, formally sided with the referees and rejected the paper stating,   to paraphrase,   that ``all the nuclear theorists working on the skyrmion model dropped the subject and are not working on it anymore. Therefore this paper cannot be accepted in PRL."  It's true that much progress is yet to be made to have the W/H Project come to fruition. And there is the RAON/IBS to come. What we have uncovered there must merely be the tip of a giant iceberg, a lot to be explored.  The multifacet of the proton structure -- a bag of quarks/gluons, a half-skyrmion-half-quark-bag at``magic-angle"~\cite{RGB} (not touched on in this note), a bound half-skyrmions etc. -- is fascinating and intriguing.  This situation is very aptly captured by the 2010 editorial in NATURE~\cite{nature}: ``After re-emerging from the depths of obscurity several times over, the spotlight is back on skyrmions.  And a reader can only wonder what other neglected gems of mathematical ideas are tucked away in the literature, awaiting a creative scientist to recognize their value to the physical world?" This remark was referring to the fantastic developments being made in condensed matter physics. We are sure that it will also be the case in the physics of compressed baryonic matter.

\end{document}